\documentclass[aps,prl,twocolumn,superscriptaddress,amsmath,showpacs
]{revtex4}
\usepackage{graphicx}
\usepackage{subfigure}

\begin{document}
\newcommand{\ket}[1]
{\left|#1\right\rangle}

\newcommand{\bra}[1]
{\left\langle #1\right|}

\newcommand{\braket}[2]
{\left\langle #1\right|\left.#2\right\rangle}

\title{High Fidelity Teleportation of Continuous Variable Quantum States using Delocalized Single Photons}

\author{Ulrik L. Andersen} 
\affiliation{Department of Physics, Technical University of Denmark, Fysikvej, 2800 Kgs. Lyngby, Denmark}

\author{Timothy C. Ralph} 
\affiliation{Centre for Quantum Computation and Communication Technology, School of Mathematics and Physics, University of Queensland, St. Lucia, Queensland 4072, Australia}

\date{\today}

\begin{abstract}
Traditional continuous variable teleportation can only approach unit fidelity in the limit of an infinite (and unphysical) amount of squeezing. We describe a new method for continuous variable teleportation that approaches unit fidelity with finite resources. The protocol is not based on squeezed states as in traditional teleportation but on an ensemble of single photon entangled states. We characterize the teleportation scheme with coherent states, Schr\"{o}dinger cat states and two-mode squeezed states and we find several situations in which near-unity teleportation fidelity can be obtained with modest resources.   
\end{abstract}


\maketitle

Quantum teleportation is the process of transmitting a quantum state through a classical channel by exploiting preshared entangled states between the sender and the receiver. Teleportation comes in two different forms depending on whether a discrete or a continuous variable (CV) is being teleported~\cite{bennett92,braunstein98,ralph98}. Examples of discrete variable versions are the teleportation of the polarization of a single photon and the spin of a single ion, whereas examples of CV teleportation include the quadratures of light beams and the collective spins of an atomic ensemble~\cite{tele1,tele2,tele3,tele4,tele5}. 
The latter type of teleportation is particularly interesting as it can be carried out determinisitically and with high efficiency using relatively simple states, transformations and detectors: Using only Gaussian squeezed states, Gaussian transformation and homodyne detectors, the teleportation of coherent states~\cite{tele3}, squeezed states~\cite{sqz}, entangled states~\cite{ent} and Schr\"{o}dinger cat states~\cite{cat} has been demonstrated. 

Despite the apparent success of these implementations, they all suffer from one major drawback: The teleportation fidelity is strongly limited and the highest fidelity measured to date with state-of-the-art technology is 83\%~\cite{highfid}. Moreover, the limited fidelity is not only a technically induced limitation; even in principle, it is not possible to reach a teleportation fidelity of 100\% using present CV teleportation schemes since such a fidelity will require infinitely squeezed, thus unphysical, resources. In other words, it is not possible - even in principle - to perform ideal CV teleportation using Gaussian squeezed state resources. 

In this Letter we propose a new scheme for the teleportation of CV states that can yield a teleportation fidelity close to 100\% with modest resources. Instead of using the standard two-mode squeezed state as a resource, we suggest to use a supply of maximally entangled single photon states in a multi-mode interferometric setting. We characterize the teleportation protocol by quantifying the performance of teleporting coherent states, cat states and two-mode squeezed states. For states or ensembles with low average energy we particularly find a strong advantage to the new approach. Maximally entangled single photon states can be generated between two parties that are connected by a lossy channel \cite{duan}, and thus the presented scheme is particularly robust against losses of the quantum channel.     

Teleportation of CVs is usually carried out by distributing two-mode squeezed states (Einstein, Podolsky and Rosen (EPR) entanglement) between two locations; the sender and the receiver~\cite{braunstein98}. A CV Bell state measurement is then carried out at the sender station 
and the resulting classical bits are used to unitarily displace the quadratures of the second half of the entangled state at the receiving station. Using this standard strategy, the fidelity of teleporting an arbitrary coherent state with unit gain is given by $F=1/(1+V)$
where $V$ is the two-mode squeezing variance of the EPR resources. In the limit of strong entanglement, $V<<1$ we obtain $F\approx1-V\approx 1-(2\bar{n})^{-1}$
%
%
where $\bar n$ is the average photon number of the entangled resource. Clearly $\bar n$ diverges as we try for very high fidelities. For example, to obtain $F \ge 0.999$ requires $\bar n \ge 500$.
 
\begin{figure}[t]
\begin{center}
\includegraphics[width=0.35\textwidth]{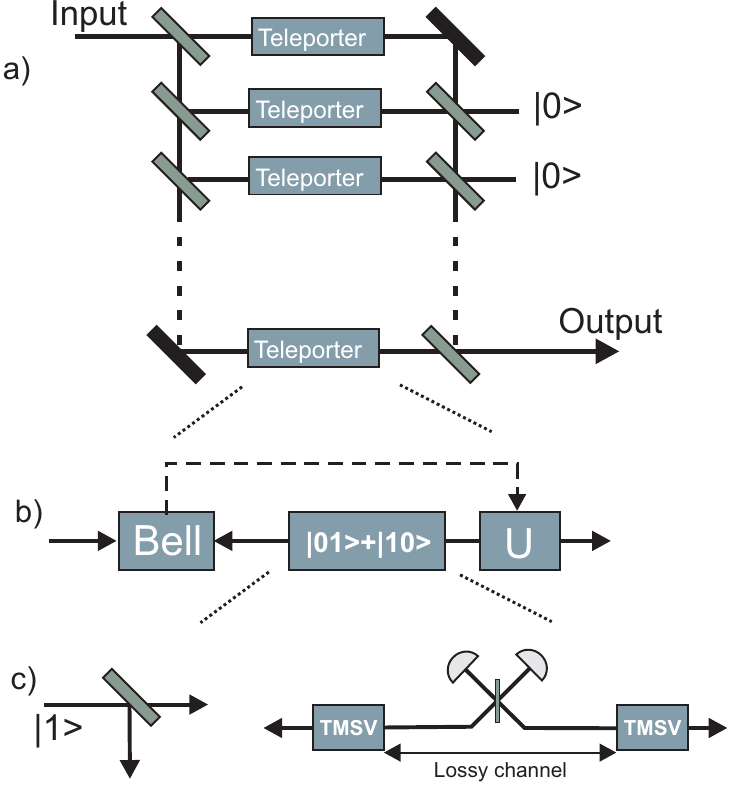}
\caption{Schematic of the proposed teleportation scheme. a) Complete teleportation scheme comprising $N$ qubit-teleporters in a multi-mode interferometer. The input state is divided into $N$ modes each of which are teleported with a qubit-teleporter. The teleported outputs are recombined in an $N$-splitter and success is obtained if all photons exist one port. b) Schematic of a qubit teleporter. A maximally entangled state is combined with a qubit in a Bell measurement, and the outcomes are used to transform the remaining mode of the entangled state. c) Generation of single photon entanglement using a single photon state or two-mode squeezed vacuum (TMSV).}
\label{schematic}
\end{center}
\end{figure}

This intrinsic limitation of CV teleportation can be overcome by using the circuit illustrated in Fig.~\ref{schematic}a. Here the non-maximally entangled two-mode squeezed state resource is replaced with maximally entangled single photon states of the form
\begin{eqnarray}
|\Phi\rangle=\frac{1}{\sqrt{2}}\left(|10\rangle+|01\rangle\right)
\label{single}
\end{eqnarray}
where $|0\rangle$ and $|1\rangle$ are the vacuum and single photon states. Such an entangled state can be easily generated by splitting a single photon on a balanced beam splitter (see Fig.~\ref{schematic}c), and it can be subsequently used to teleport an arbitrary state in a two-dimensional Hilbert space spanned by the vacuum and the single photon state~\cite{martini,lvovsky} (see Fig.~\ref{schematic}b). Due to the low dimensionality of this Hilbert space, maximally entangled states can be produced in practice and the teleportation fidelity can in principle be as high as 100\%. However, since the teleported state is limited to two dimensions, an input state with a higher dimension will be truncated; only the vacuum and single photon terms will survive in the teleporter~\cite{pegg,lvovsky}. To circumvent truncation of the input state we suggest to use an ensemble of the single photon entangled states in (\ref{single}), and use them to execute a larger number of qubit teleportations in a multi-mode interferometer as illustrated in Fig.~\ref{schematic}a: The input state to be teleported is divided evenly between $N$ modes using an array of  $N-1$ beam splitters (an $N$-splitter) such that each mode contains much less than a single photon on average. Each one of the modes in the interferometer can then be efficiently teleported using the single photon resources. After the $N$ teleportations, the $N$ modes are recombined interferometrically using another $N$-splitter to produce $N$ output modes. The final teleported state is heralded in the original mode when no photons are registered in the remaining modes. If $N$ is large compared to the average photon number of the input state, the probability of any photons exiting the remaining ports will be very small, and thus the efficiency (or even the presence) of the heralding detectors is not critical.
We note that the idea of splitting a coherent state for individual processing of two-dimensional sub-spaces has also been used for noiseless amplification~\cite{ralph,pryde} and optimal coherent state discrimination~\cite{Silva12}.

We start by considering the teleportation of a coherent state $|\alpha\rangle$, using the circuit outlined above. The $N$-splitter transforms the input coherent state into a product state of $N$ coherent states with reduced amplitude: $|\beta\rangle^{\otimes N}=|\alpha/\sqrt{N}\rangle^{\otimes N}$.
These states are then combined with the entangled resources to form $|\beta\rangle^{\otimes N}\otimes|\Phi\rangle^{\otimes N}$ and teleported using one of the Bell projectors and conditional unitary transformations.
This will truncate the Hilbert spaces of the individual modes to 2 dimensions and the teleported states can be cast in the form
\begin{eqnarray}
|\beta\rangle^{\otimes N} \rightarrow \left(\exp(-|\alpha|^2/(2N))\left(1+\frac{\alpha}{\sqrt{N}}\hat{a}^\dagger\right)|0\rangle\right)^{\otimes N}
\end{eqnarray}
After teleportation, the states will be coherently recombined on $N$ beam splitters by projecting all outputs onto the vacuum state except one. The final output state is      
\begin{equation}
|\phi\rangle=\frac{e^{-|\alpha|^2/2}}{\sqrt{P_{suc}}}\sum_{k=0}^{N}\binom{N}{k}\left(\frac{\alpha}{N}\right)^k\sqrt{k!}|k\rangle
\end{equation}
which approaches the input state for large $N$, and the success rate is
\begin{equation}
P_{suc}=e^{-|\alpha|^2}\sum_{k=0}^N\binom{N}{k}^2\left(\frac{|\alpha|^2}{N^2}\right)^kk!
\end{equation}
(See supplemental information for details about the derivation.)
The fidelity in teleporting an ensemble of coherent states will be investigated below when teleportation of one mode of an EPR state is treated, but first we generalize the protocol to include the teleportation of an arbitrary state. 
We note from the analysis above that the Fock state is transformed through the teleporter as  
$|k\rangle\rightarrow\binom{N}{k}\frac{k!}{N^k}|k\rangle$.
An arbitrary pure input state $|\psi\rangle =\sum_{k=0}^\infty{c_k|k\rangle}$
is therefore transformed into the pure output state 
\begin{equation}
|\psi\rangle_{tele} =\kappa\sum_{k=0}^N{c_k\binom{N}{k}\frac{k!}{N^k}|k\rangle}
\end{equation} 
where $\kappa$ is a normalization constant.

\begin{figure}[t]
\begin{center}
\includegraphics[width=0.5\textwidth]{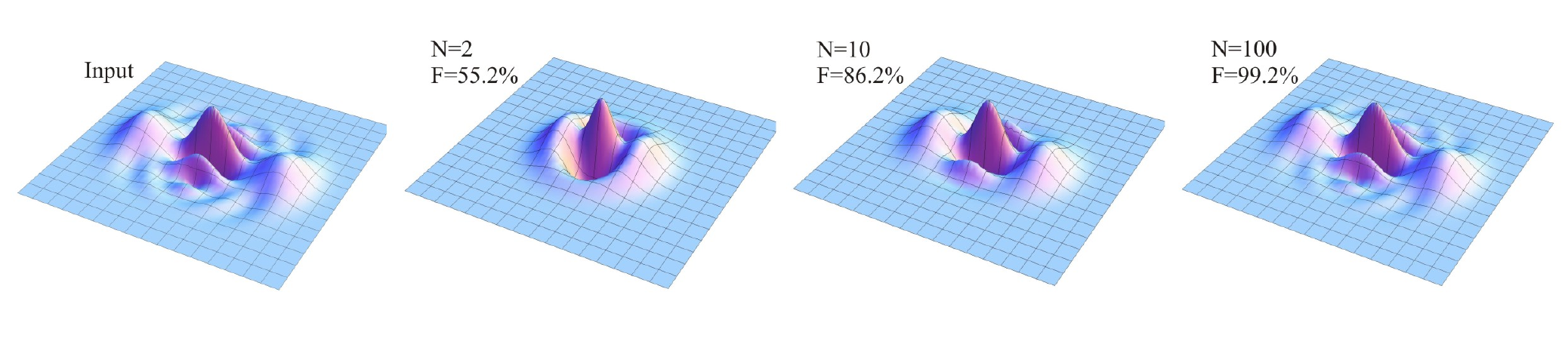}
\caption{Wigner functions of input and teleported cat states.}
\label{catfig}
\end{center}
\end{figure}

To illustrate the teleportation protocol with non-Gaussian states we consider the teleportation of a coherent state superposition; $(|\alpha\rangle+|-\alpha\rangle)/\sqrt{2+2 \exp(-2\alpha^2)}$ where $\alpha$ is the amplitude. The Wigner functions of the input and teleported output states are presented in Fig.~\ref{catfig} for $\alpha=2$ and various values of $N$. It is seen that by using 100 single photon states, the teleportation fidelity is as high as 99.2\%. In contrast, the fidelity for teleporting a coherent superposition of this kind using the standard approach 
%
leads to the conclusion that $V=0.001$ (i.e. $\bar n = 500$ or 30dB squeezing) is required to obtain a similar fidelity~\cite{braunstein98}.  

We now consider the teleportation of one mode of a two-mode squeezed (or EPR) state via our protocol. There are two reasons why this is an important case to study. Firstly, the teleportation of a single mode of an EPR state models the teleportation of a Gaussian ensemble of different pure states such as a coherent state ensemble or a Fock state ensemble. The constituent states of the ensemble are determined by the type of measurement made on the mode which is not teleported, and the width of the distribution is given by the degree of two-mode squeezing. For example, heterodyne measurement will produce an ensemble of coherent states, whilst number measurements will produce an ensemble of Fock states. The second reason for studying the teleportation of the EPR state is that it is a type of entanglement swapping that can be used to distribute entanglement between locations that have never directly coherently interacted. 

The EPR state can be written as 
\begin{eqnarray}
|EPR\rangle =\sqrt{1-\chi ^2}\sum_{k=0}^{\infty}\chi^k|k\rangle|k\rangle
\label{EPR}
\end{eqnarray} 
where the strength of entanglement is given by the parameter $\chi$ with $\chi =0$ and $\chi=1$ corresponding to no and maximal entanglement, respectively. The strength of the entanglement is directly related to the degree of two-mode squeezing through the relation $V_s=(1-\chi)/(1+\chi)$ where $V_s=\langle(x_A+x_B)^2\rangle=\langle (p_A-p_B)^2\rangle$ and $x(p)$ are the amplitude (phase) quadratures of the mode $A$ and $B$~\cite{reid}. If we consider the reduced state of, say mode $B$, we find that it is the thermal state (or an ensemble of coherent states) with variance $1/2(V_s + 1/V_s)$.

By transforming mode $B$ through our teleportation protocol, the output is found to be
\begin{eqnarray}
|EPR\rangle_{tele} =\frac{1}{\sqrt{P_{suc}}}\sqrt{1-\chi ^2}\sum_{k=0}^{\infty}\chi^k\binom{N}{k}\frac{k!}{N^k}|k\rangle|k\rangle
\label{teleEPR}
\end{eqnarray} 
where the success rate (assuming deterministic Bell measurements) is
\begin{eqnarray}
P_{suc}=(1-\chi ^2)\sum_{k=0}^{\infty}\chi^{2k}\binom{N}{k}^2\frac{k!^2}{N^{2k}}
\end{eqnarray} 
The quality of the teleporter can be quantified in different ways. Considering the teleportation of an ensemble of pure states, the fidelity is an appropriate parameter, while the teleportation of entanglement (or entanglement swapping) is usefully characterized by the degree of two-mode squeezing of the resulting state. We first consider the fidelity, which is the overlap squared between the states in (\ref{EPR}) and (\ref{teleEPR}):
\begin{eqnarray}
F_{EPR} = ({{(1 - \chi^2)} \over {\sqrt{P_{suc}}}} \sum^N_{k = 
     0} \chi^{2  k}\binom{N}{k} {{ k!}\over {N^k}} )^2
\label{FEPR1}
\end{eqnarray}  
However, this fidelity is not identical to the averaged fidelity of the constituent states of the ensemble modelled by the EPR state. On the other hand, it
is straightforward to show that $F_{EPR}$ is equivalent to the square of the average amplitude fidelity of the constituent states of the ensemble. Explicitly, $ F_{EPR} = ({{1}\over{K}} \sum^K_{i=1} \sqrt{F_i})^2$
where $F_i$ is the fidelity of teleportation for the $i$th constituent state of the ensemble (and the sum goes to an integral in the limit of an infinite number of constituent states). This expression, along with the positivity of the variance (i.e. $\bar{x^2} -{\bar x}^2 \ge 0$), can then be used to place a lower bound on the average fidelity of teleportation for the ensemble: 
$F_{EPR} \le \bar F_e = {{1}\over{K}} \sum^K_{i=1} F_i$ \cite{nielsen}.
Therefore, the fidelity computed in (\ref{FEPR1}) sets a lower bound on teleporting a Gaussian ensemble of pure states. Some values for $F_{EPR}$ with associated variances, $V_s$, photon number resources, $N$, and success probability, $P_{suc}$, are found in Table I. For comparison, we also insert the mean photon numbers of a two-mode squeezed state, $\bar{n}$, that is required to attain similar performance for an optimised standard CV teleportation protocol~\cite{MAR03,RAL99}.          
%
%
%
%
%
%
\begin{table}[htdp]\begin{center}\begin{tabular}{c|c|c|c|c}$1/V_s$ & 2 & 3 & 5 & 7  \\\hline $F_{EPR}$ & 0.99 & 0.99 & 0.99 & 0.91 \\\hline $N$ & 1 & 4 & 17 & 6  \\\hline $\bar n$ & 17 & 50 & 111 & 17 \\\hline $P_{suc}$ & 0.99 & 0.97 & 0.95 & 0.80 \end{tabular} \caption{Comparison of the single photon resources, $N$, required by our new teleportation protocol versus the average photon number, $\bar n$, required in the entanglement source for the standard protocol, for achieving the same fidelity of teleportation of one arm of an EPR of squeezing $V_s$. $V_s$ can either be interpreted as the strength of the correlation in the entanglement, or as the variance of an ensemble of states,  $V_{ens} = 1/2(V_s + 1/ V_s)$, being teleported. The probability of success of the new protocol, $P_{suc}$, is also shown}\end{center}\label{defaulttable}\end{table}

\begin{figure}[t]
\begin{center}
\includegraphics[width=0.4\textwidth]{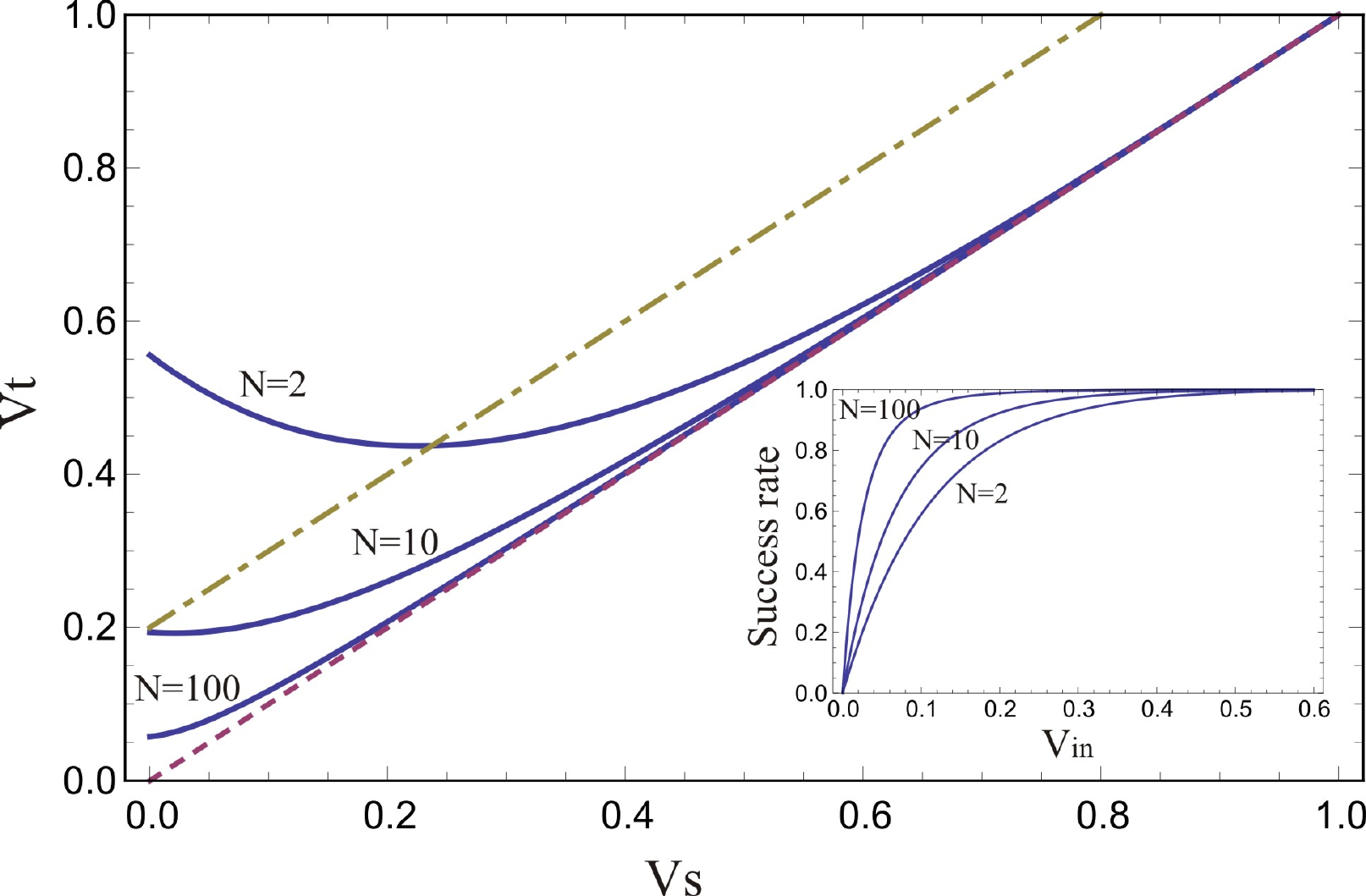}
\caption{Two-mode squeezing variance of the teleported state as a function of the input state for different resources (see text). The inset is the success rate assuming ideal Bell measurements. The dot-dashed line shows the performance of a standard CV teleporter with $10$dB squeezing. The dashed curve is the performance of a hypothetical perfect CV teleporter with infinite squeezing.}
\label{teleTMS}
\end{center}
\end{figure}

As mentioned above, for the purpose of distributing CV entanglement it is appropriate to quantify the teleportation performance of the EPR state by determining the two-mode squeezing variances $V_{t}$ of the teleported state and comparing this to the variance of the input state, $V_s$. We note that since the output state is not necessarily a Gaussian state (it only approaches a Gaussian state for large $N$), the two-mode squeezing variance is not an exact measure of entanglement but it is known to be a lower bound and it yields the amount of entanglement useful for Gaussian operations. Due to the phase space symmetry of the teleporter, we have that $\langle(x_A+x_B)^2\rangle_t =\langle(p_A-p_B)^2\rangle_t=V_t$ and   
\begin{eqnarray}
V_{t}
&=& \frac{1}{1-\chi^{(2)}}\sum_{k=0}^{\infty}[\chi^{2k}\binom{N}{k}^2\frac{k!^2}{N^{2k}}(1+2k) \nonumber \\
&& +2\chi^{2k+1}\binom{N}{k+1}\binom{N}{k}\frac{(k+1)!k!}{N^{2k+1}}(1+k)]
\end{eqnarray} 
The variance of the teleported state is plotted in Fig.~\ref{teleTMS} against the variance of the input state for three different numbers of photon resources ($N=2,10,100$). For comparison we also evaluate the performance of a standard CV teleportation protocol of an EPR state (known as CV entanglement swapping~\cite{CVswap}) assuming 10dB of squeezing of the resource. The result is plotted in Fig.~\ref{teleTMS} by the dot-dashed curve, and we clearly see that for most EPR input states, the new teleporter outperforms the standard entanglement swapping protocol. 
The success rate assuming ideal Bell measurements is shown at the inset of Fig.~\ref{teleTMS}.


The teleportation protocol developed in this paper is only deterministic when the number of single photon resources is much larger than the average number of photons of the input state (such that all photons exit a single port of the interferometer in Fig.~\ref{schematic}a) and if the Bell state measurement is conducted with unit efficiency. Both of these requirements can be difficult to meet with current technology (see below for a discussion). However, non-unit probability of success leads to locatable or heralded errors, i.e. one knows when the error has occurred because the protocol fails. This is in distinction with non-unit fidelity which indicates the occurrence of non-locatable errors. In many quantum information tasks, such as the distribution of entanglement, heralded errors are far preferable to non-locatable errors. Thus the achievement of high fidelities, even if accompanied by finite probability of success, can be advantageous (see Table I).

Another advantage of using the outlined scheme to teleport EPR entanglement is that it can be done with high fidelity even if the quantum transmission channel is lossy. This stems from the fact that maximally entangled single photon entanglement can be prepared between distant sites that are connected by lossy channels~\cite{duan,brask}. The idea is to use local sources of two-mode squeezed vacua produced e.g. by parametric down conversion as illustrated in Fig.~\ref{schematic}c. One mode from each site is sent through a lossy quantum channel and half way through the channel the resulting two modes interfere at a symmetric beam splitter. The two outputs are detected with single photon counters and the measurement of a single photon heralds the state (residing at the sender and receiving station) in an entangled state of the kind in (\ref{single}) independent of the channel and detector losses. However, the channel loss limits the success rate at which the entangled states can be prepared. To increase the rate (or distance for a given rate), entanglement swapping of single photon entangled states must be applied~\cite{brask}. See also supplemental information for further discussions on the effect of inefficient single photon generation. 

In order to carry out deterministic teleportation, all four Bell projectors must be implemented. Two of
them (the singlet, $|\Phi_-\rangle=\frac{1}{\sqrt{2}}\left[|0,1\rangle-|1,0\rangle\right]$, and the triplet projector, $|\Phi_+\rangle=\frac{1}{\sqrt{2}}\left[|0,1\rangle+|1,0\rangle\right]$) are simple to implement with linear optics.
However, the two other Bell states cannot be easily identified thus the probability of success is limited to 50\%. Since this would be the success rate for each qubit teleporter of the interferometer, the overall success rate would scale as $(1/2)^N$. 
To implement all projectors deterministically, single photon auxiliary states~\cite{grice11} or in-line nonlinear optics~\cite{bjork11} are required. Several proposals on the feasible implementation of all Bell measurements using weak non-linearities (in the so-called Purcell regime) have been put forward~\cite{Bonato2010,Witthaut2012}, and current experimental progress in the scalable solid state regime is remarkably fast. Strong coupling of Nitrogen-Vacancy centers in diamond to nano-wires~\cite{Harvard,Huck} as well as coupling of quantum dots to nano-cavities~\cite{Englund2007,gazzano} have been realized with a high Purcell effect. Such devises can be used both as deterministic single photon sources as well as for the implementation of deterministic Bell measurements, and due to the intrinsic scalability of solid state emitters, a near-deterministic and near-unit fidelity implementation of the suggested CV teleportation protocol maybe within reach in the next few years. A teleportation scheme with $N$=2 and linear Bell measurements can be readily implemented using a pair of down-converters based on the setup in Ref.~\cite{Pryde}.    

In summary, we have proposed a scheme for the teleportation of CVs using a supply of single photon states. This scheme allows for near-unit fidelity in the teleportation of arbitrary states in a CV Hilbert space with the use of finite resources. We have in particular investigated the teleportation of a cat state of light as well as one mode of an entangled state, and in both cases we find very high fidelities using modest resources. Our new scheme has clear advantages over the standard approach when the states to be teleported have low average energy. Although the technology needed to realize the scheme near deterministically is still under development, non-deterministic realizations in useful scenarios are practical today using linear optical techniques.
We finally note that since the CV teleportation protocol is a fundamental element in many CV quantum information protocols such as quantum repeaters, measurement-induced quantum computation and quantum error correcting codes, the proposed high-fidelity teleportation strategy will impact many areas of CV quantum information processing.

We acknowledge support from the Danish Council for Independent Research (Technology and Production Sciences and Natural Sciences) and the Australian Research Council (Project number CE110001027). TCR acknowledges discussions with Geoff Pryde.




 
 


\end{document}